\documentclass[prl, superscriptaddress,twocolumn, amsmath, amssymb, showpacs]{revtex4-2}
\usepackage{CJK}
\usepackage{color}
\usepackage{sidecap}
\usepackage{times}
\usepackage{verbatim}
\usepackage{graphicx}
\usepackage{color}
\usepackage{graphics}
\usepackage{amsmath}
\usepackage{float}
\usepackage{ulem}
\usepackage{lineno}

\usepackage{amssymb}
\usepackage[colorlinks,linkcolor=blue,
            citecolor=blue,
            hyperindex,
            pdfstartview=FitH,
            plainpages=false]
            {hyperref}

\newcommand {\mJcm}{mJ/cm$^2$}
\newcommand {\dxz}{$d_{xz}$}
\newcommand {\dyz}{$d_{yz}$}
\newcommand {\dz}{$d_{z^2}$}
\newcommand {\fco}{$F_{c1}$}
\newcommand {\fct}{$F_{c2}$}

\begin{document}
\title{Anomalous contribution to the nematic electronic states from the structural transition in FeSe revealed by time- and angle-resolved photoemission spectroscopy}
\author{Yuanyuan Yang}
\affiliation{Key Laboratory of Artificial Structures and Quantum Control (Ministry of Education), Shenyang National Laboratory for Materials Science, School of Physics and Astronomy, Shanghai Jiao Tong University, Shanghai 200240, China}
\author{Qisi Wang}
\affiliation{State Key Laboratory of Surface Physics and Department of Physics, Fudan University, Shanghai 200433, China}
\author{Shaofeng Duan}
\affiliation{Key Laboratory of Artificial Structures and Quantum Control (Ministry of Education), Shenyang National Laboratory for Materials Science, School of Physics and Astronomy, Shanghai Jiao Tong University, Shanghai 200240, China}
\author{Hongliang Wo}
\affiliation{State Key Laboratory of Surface Physics and Department of Physics, Fudan University, Shanghai 200433, China}
\author{Chaozhi Huang}
\author{Shichong Wang}
\author{Lingxiao Gu}
\affiliation{Key Laboratory of Artificial Structures and Quantum Control (Ministry of Education), Shenyang National Laboratory for Materials Science, School of Physics and Astronomy, Shanghai Jiao Tong University, Shanghai 200240, China}
\author{Dao Xiang}
\affiliation{Key Laboratory for Laser Plasmas (Ministry of Education), School of Physics and Astronomy, Shanghai Jiao Tong University, Shanghai 200240, China}
\affiliation{Tsung-Dao Lee Institute, Shanghai Jiao Tong University, Shanghai 200240, China}
\author{Dong Qian}

\affiliation{Key Laboratory of Artificial Structures and Quantum Control (Ministry of Education), Shenyang National Laboratory for Materials Science, School of Physics and Astronomy, Shanghai Jiao Tong University, Shanghai 200240, China}
\affiliation{Tsung-Dao Lee Institute, Shanghai Jiao Tong University, Shanghai 200240, China}
\affiliation{Collaborative Innovation Center of Advanced Microstructures, Nanjing University, Nanjing 210093, China}
\author{Jun Zhao}
\affiliation{State Key Laboratory of Surface Physics and Department of Physics, Fudan University, Shanghai 200433, China}
\affiliation{Institute of Nanoelectronics and Quantum Computing, Fudan University, Shanghai 200433, China}
\author{Wentao Zhang}
\email{wentaozhang@sjtu.edu.cn}
\affiliation{Key Laboratory of Artificial Structures and Quantum Control (Ministry of Education), Shenyang National Laboratory for Materials Science, School of Physics and Astronomy, Shanghai Jiao Tong University, Shanghai 200240, China}
\affiliation{Tsung-Dao Lee Institute, Shanghai Jiao Tong University, Shanghai 200240, China}
\affiliation{Collaborative Innovation Center of Advanced Microstructures, Nanjing University, Nanjing 210093, China}

\date {\today}

\begin{abstract}

High-resolution time- and angle-resolved photoemission measurements were made on FeSe superconductors. With ultrafast photoexcitation, two critical excitation fluences that correspond to two ultrafast electronic phase transitions were found only in the $d_{yz}$-orbit-derived band near the Brillouin-zone center within our time and energy resolution. Upon comparison to the detailed temperature dependent measurements, we conclude that there are two equilibrium electronic phase transitions (at approximately 90 and 120 K) above the superconducting transition temperature, and an anomalous contribution on the scale of 10 meV to the nematic states from the structural transition is experimentally determined. Our observations strongly suggest that the electronic phase transition at 120 K must be taken into account in the energy band development of FeSe, and, furthermore, the contribution of the structural transition plays an important role in the nematic phase of iron-based high-temperature superconductors.

\end{abstract}

\maketitle
Nematic order in iron-based superconductors refers to an electronic phase with broken rotational symmetry but preserved translational symmetry, and usually occurs at decreasing temperature combined with a lattice structure transition from tetragonal to orthorhombic phase. Such a nematic phase may play a pivotal role in high-temperature superconducting pairing, and its underlying physics is a predominant topic in the study of iron-based systems \cite{Paglione2010,Chuang2010,Bohmer2015,Yi2011,Fu2012,Lu2014}. 
The electronic nematic phase can be a result of the tetragonal-to-orthorhombic structural phase transition, orbital correlation, or the set-in of magnetic order, and early theoretical and experimental work suggested that the tetragonal-to-orthorhombic lattice distortion may be too small to drive the nematic electronic properties, such as in-plane resistivity \cite{Chandra1990,Chu2010,Tanatar2010, Fernandes2014,Kothapalli2016,Tanatar2016}. However, there is still a lack of evidence of a purely electronic (nematic) phase transition, and how much the structural transition contributes to the band renormalization is still not clear experimentally. 

FeSe, with a superconducting transition at approximately 9 K, undergoes a tetragonal-to-orthorhombic lattice deformation and a breaking of rotational symmetry of electronic order (nematic phase) at the same temperature of approximately 90 K without the long-range magnetic order as discovered in many other iron-based materials, allowing us to study the pure nematic phase over a wide temperature range \cite{Hsu2008,Margadonna2008,McQueen2009,Bendele2010,Bohmer2015,Baek2015,Boehmer2013}. However, the extent of the individual contributions of the structural transition and the purely electrical order to the nematic electronic states is still unknown. As most characteristics of nematic order are measured at thermal equilibrium conditions \cite{Nakayama2014,Borisenko2016,Suzuki2015,Zhang2015,Watson2016,Fedorov2016,Watson2017,Pfau2019}, it is difficult to differentiate the nematic electronic order from that of the structural transition and determine the contribution to the nematic states from the structural transition. Taking advantage of ultrafast experiments by using strong ultrafast photon pulses to excite only the electronic states, retaining the preserved orthorhombic lattice structure, it is possible to drive only the electronic phase transitions in an ultrafast manner without deforming the crystal lattice  \cite{Tang2020,Rohwer2011}. Thus, it is possible to isolate the electronic states from the lattice, and to clarify the origin of both the nematic and structural phase transitions and estimate the contribution of the structural transition to the nematicity.  

In this Letter, we report two ultrafast purely electronic phase transitions and two equilibrium phase transitions above the superconducting transition temperature in FeSe revealed by time- and angle-resolved photoemission spectroscopy (trARPES). In particular, with improved energy and pump fluence resolution \cite{Duan2021}, we demonstrate that the response of the electronic structure at several hundred femtoseconds after ultrafast photoexcitation exhibits a three-excitation-regime behavior with two critical excitation fluences of $F_{c1}$ $\approx$ 0.06 and $F_{c2}$ $\approx$ 0.2 \mJcm~only in the \dyz-orbit-derived band within resolution, consistent with two equilibrium phase transitions found in the temperature dependent measurement.
It is quite anomalous that the \dxz~and \dz~ bands shift downwards below \fco~after ultrafast photoexcitation, while in contrast they shift upwards in the equilibrium temperature dependent measurement below the structural transition temperature.
By comparing the non-equilibrium and equilibrium measurements, we conclude that the nematic phase near the zone center of FeSe is closely related to the \dyz~orbit and the contribution to the nematicity from the structural transition cannot be negligible but comparable to that from the electronic transition.

\begin{figure*}
\centering\includegraphics[width=2\columnwidth]{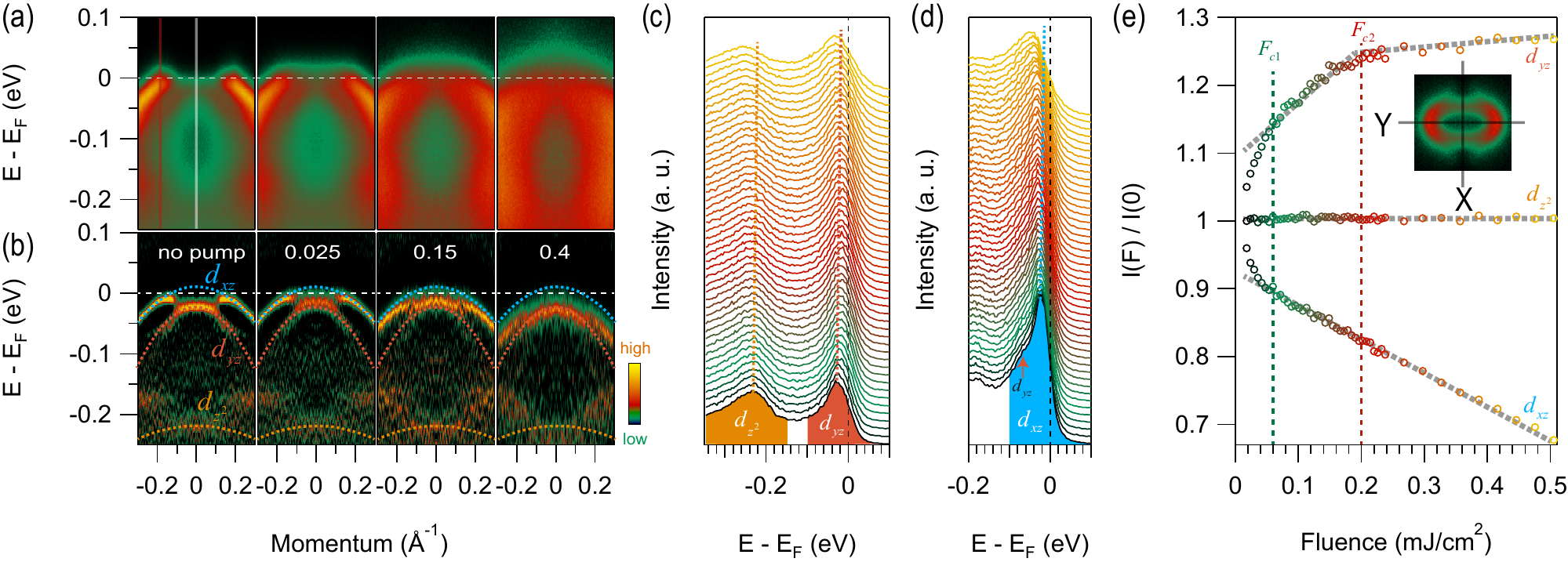}
\caption{
Fluence-dependent trARPES spectra measured along Y near the Brillouin-zone center at an equilibrium temperature of 4.5 K.
(a) TrARPES spectrum at equilibrium ($-$0.1 ps), shortly after photoexcitation (0.2 ps) with pump fluences of 0.025, 0.15, and 0.4 \mJcm.
(b) Corresponding second-derivative images from (a). Blue, red, and orange dashed lines are the guides to show the equilibrium bands of $d_{xz}$, $d_{yz}$, and $d_{z^2}$.
(c) Fluence-dependent energy distribution curve (EDC) at $\Gamma$ (gray cut shown in left-hand panel of (a)). 
(d) Fluence-dependent EDC at Fermi momentum (red cut shown in left-hand panel of (a)). The fluence of each EDC is represented by the corresponding color shown in (e).
(e) Integrated band intensities from (c) and (d) as function of pump fluence for three measured bands.
}
\label{Fig1}
\end{figure*}

In the trARPES measurements, infrared pump laser pulses with a photon energy $h\nu$ = 1.77 eV and repetition rate of 500 kHz drive the sample into non-equilibrium states \cite{Yang2019}, and ultraviolet probe pulses (6.05 eV, s-polarized) subsequently photoemit electrons. The energy and time resolution are 16.3 meV and 113 fs respectively, giving a  time-bandwidth product of approximately 1840 meV$\cdot$fs, close to the physical limit for Gaussion pulses. The spot sizes of the pump and probe beam are about 100 and 12 $\mu m$, respectively, giving a pump fluence resolution of about 2.5\%\cite{Duan2021}. Thermal drift of the sample position was automatically corrected by a computer with a precision less than 1 $\mu m$, ensuring the measurements were took on a fixed spot with a precision of less than 1 $\mu m$. FeSe single crystals were grown using KCl-ACl$_3$ flux under a permanent temperature gradient\cite{Wang2016}, and cleaved under ultrahigh-vacuum conditions with a pressure of 3$\times$10$^{-11}$ Torr. All trARPES measurements were taken at an equilibrium temperature of approximately 4.5 K.

Equilibrium photoemission spectra probed by the 6.05-eV laser demonstrate three bands \cite{Watson2015,Gerber2017}, including one hole-like band ($d_{xz}$) crossing the Fermi level, another band ($d_{yz}$) with its top at approximately -20 meV, and one flat band ($d_{z^2}$) located at -0.2 eV (Figs.~\ref{Fig1}(a) and (b)). With a low pump fluence of 0.025 \mJcm, photoemission spectra at 0.2 ps after photoexcitation show an up-shift of the $d_{yz}$ and a down-shift of the $d_{xz}$ band, which are more remarkable upon enhancing the pump fluence up to 0.15 \mJcm. At the highest pump fluence of 0.4 \mJcm~that was measured, the $d_{xz}$ band shifts downwards further, but the $d_{yz}$ band shifts to higher binding energy, opposite to that at low pump fluences. The aforementioned observation can be directly evidenced in the fluence-dependent energy distribution curves (EDCs) at corresponding momentum (\dyz~in Fig.~\ref{Fig1}(c) and \dxz~in Fig.~\ref{Fig1}(d)).
Around a pump fluence of approximately 0.06 \mJcm~(\fco), there are apparent changes of the slope of the photoemission intensity as a function of pump fluence for both the \dxz~and \dyz~ bands (Fig.~\ref{Fig1}(e)). Above a pump fluence of approximately 0.2 \mJcm~(\fct), the photoemission intensity of the \dyz~band is almost unchanged upon enhancing the pump fluence, but shows no resolvable change of the slope of the curve in the \dxz~band. Both the \fco~ and \fct~	features are absent in the \dz~band.
We attribute the \fco~feature to the critical pump fluence that destroys the electronic nematic order and the \fct~feature to another electronic phase transition as discussed below.

\begin{figure}
\centering\includegraphics[width=1\columnwidth]{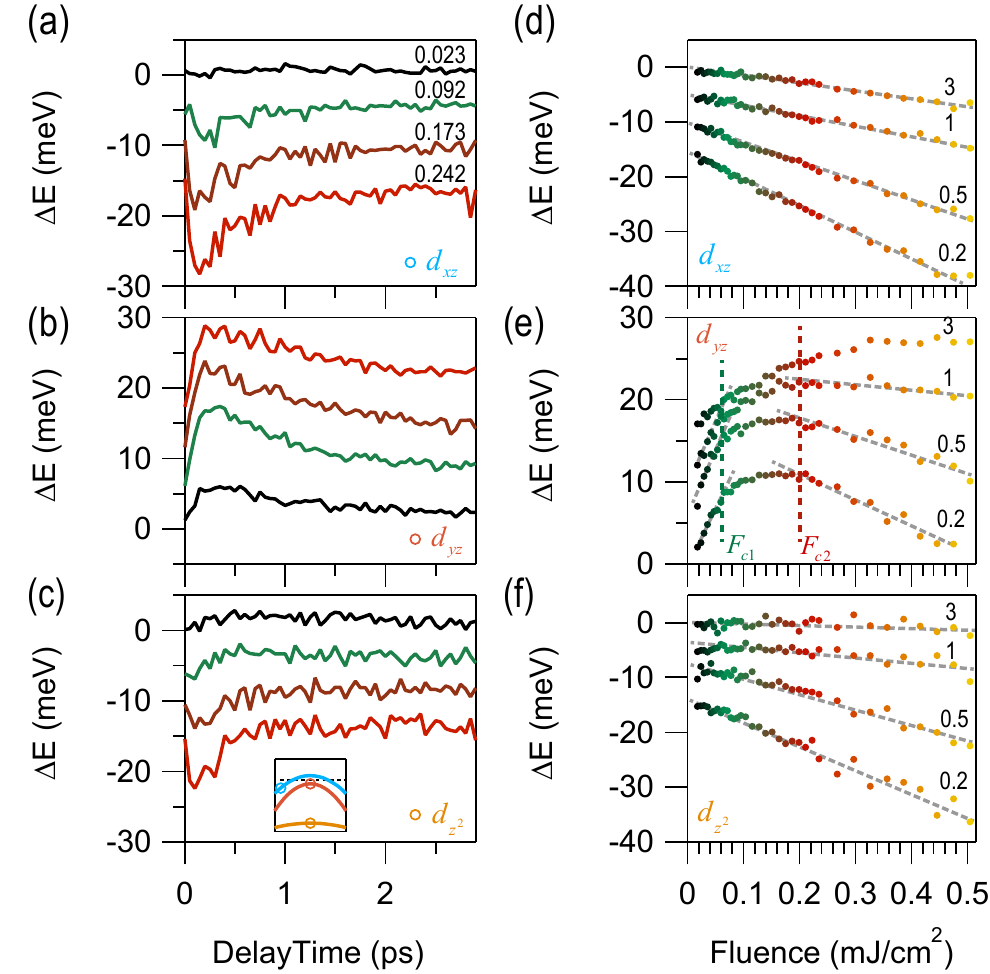}
\caption{
Time- and fluence-dependent energy shifts of $d_{xz}$, $d_{yz}$, and $d_{z^2}$ bands.
(a), (b), and (c) Time-delay-dependent energy shifts of the three aforementioned bands at pump fluences of 0.023, 0.092, 0.173, and 0.242 \mJcm. Energies of the bands are determined at the momentum shown in the inset of (c).
(d), (e), and (f) Fluence-dependent energy shifts of the three aforementioned bands at delay times of 0.2, 0.5, 1, and 3 ps. Fluence dependent measurements were taken at the fixed delay time by finely tuning the pump pulse energy. The momentum of the EDC in the analysis of the \dxz~band is at 0.25 {\AA}$^{-1}$, which is away from the Fermi momentum, to avoid the thermal broadening of the Fermi distribution.
}
\label{Fig2}
\end{figure}

The observed critical pump fluences at \fco~and \fct~are also evidenced in the ultrafast energy-band shifts (Fig.~\ref{Fig2}). The \dxz~band moves down gradually with increasing pump fluence (Fig.~\ref{Fig2}(a)), with a shift of approximately 13 meV at a pump fluence of 0.242 \mJcm~and delay time near 0.2 ps. 
In contrast, the \dyz~band up-shifts more greatly than the \dxz~band at low pump fluence (Fig.~\ref{Fig2}(b)). For the band \dz, the shift is only resolvable near time zero (Fig.~\ref{Fig2}(c)). On the basis of band shifts as a function of delay time, the band shift oscillates at a frequency of approximately 5.28 THz, which is the photoinduced $A_{1g}$ coherent phonon mode \cite{Gerber2017}. Detailed pump-fluence dependencies of the band shifts show that both the \dxz~and \dz~bands shift almost linearly to the pump fluence within our time and energy resolution (Figs.~\ref{Fig2}(d)--(f)), and only in the \dyz~band the pump-induced band shift increase significantly below the critical pump fluence \fco~ and drop above the critical pump fluence \fct~for delay times of 0.2, 0.5, and 1 ps. The absence of the two critical fluences at a delay time of 3 ps in band \dyz~suggests that the recovery time of the two phase transitions is shorter than 3 ps. We note that even under strongly non-equilibrium situation the nematic order is destroyed, but with probe beam spot comparable to the size of the single orthorhombic domain, the \dxz~and \dyz~orbits are still not equivalent, since the lattice is holding its low-temperature anisotropic structure. 

\begin{figure}
\centering\includegraphics[width=0.9\columnwidth]{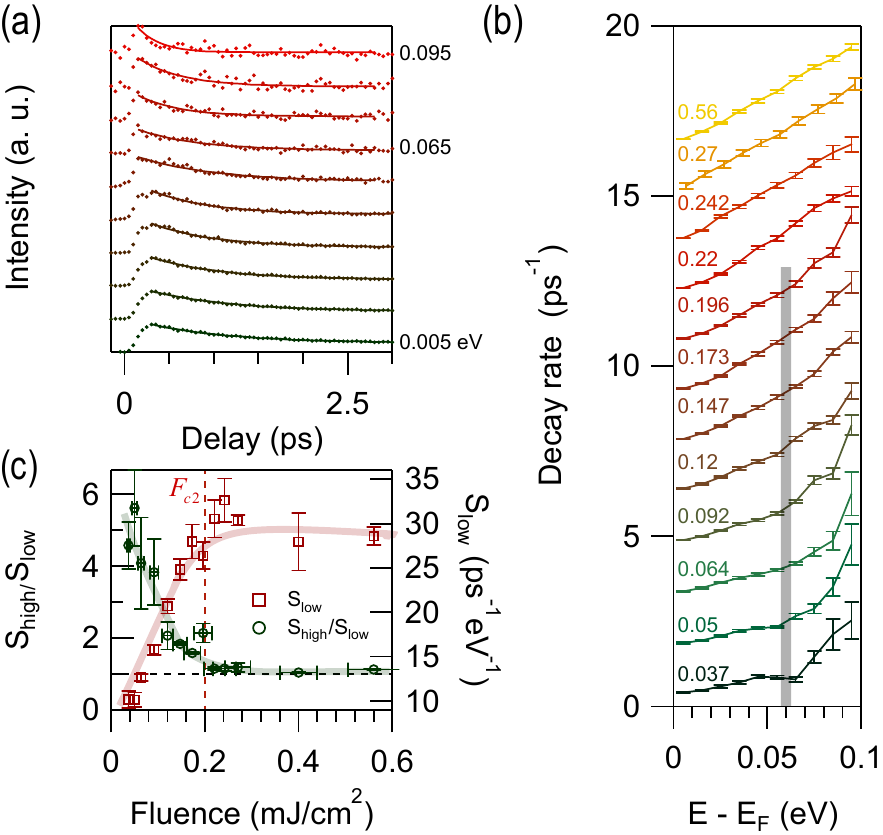}
\caption{
Non-equilibrium quasiparticle dynamics.
(a) Intensity of non-equilibrium electrons as a function of delay time from 0.005 to 0.095 eV above the Fermi energy measured at a pump fluence of 0.05 \mJcm. Solid curves are fittings to an exponential decay function $C\cdot e^{-t\gamma}$, in which $\gamma$ is the non-equilibrium quasiparticle recovery rate and $C$ is the amplitude.
(b) Non-equilibrium electron recovery rate as a function of energy for pump fluences between 0.037 and 0.56 \mJcm. 
(c) Slope of decay rate ($S_{slow}$) and ratio of $S_{high}/S_{slow}$ as a function of pump fluence. The $S_{slow}$ and $S_{high}$ are determined by fitting the decay rates below 0.05 eV and above 0.07 eV in (b) to linear functions.
}
\label{Fig3}
\end{figure}

The critical pump fluence \fct~can be also evidenced in the ultrafast evolution of the pump-induced non-equilibrium quasiparticles. It is clear that at low energy the recovery rate is much smaller than that at high energy, presenting a kink at approximately 0.06 eV (Fig.~\ref{Fig3}(a)), with the extracted quasiparticle recovery rate as a function of energy shown in Fig.~\ref{Fig3}(b). Upon enhancing the pump fluence, such a kink is becoming weaker and above approximately 0.2 \mJcm~(\fct) the decay rate is even linear to the energy. Such a critical pump fluence at \fct~is clearly evidenced by analysis of the slope of the decay rate as a function of energy in detail (Fig.~\ref{Fig3}(c)).

Such an energy scale at approximately 0.06 eV shown in the non-equilibrium quasiparticle recovery rate is an evidence of electrons scattered with some collective excitation, or a band gap at this energy. However, in FeSe the cutoff energy of the phonon modes is approximately 40 meV \cite{Zakeri2017} and there is no report of spin fluctuation or other bosonic modes at approximately 0.06 eV. Thus, such an energy scale of 0.06 eV possibly results from electrons scattered with a band gap at this energy scale near the Fermi energy. There is no such band gap near the Brillouin zone center, but instead, equilibrium ARPES experiments at the M point in the FeSe family evidenced a band gap between \dxz~and \dyz~of approximately 50 meV, which can be the orbital order parameter, below approximately 120 K \cite{Nakayama2014,Watson2015,Zhang2015,Yi2019}.
We speculate that the direct hopping of electrons between the gapped bands would excite electrons with energy below the energy gap, hindering the recovery of the low-energy non-equilibrium quasiparticles, like the role played by bosonic modes at this energy \cite{Smallwood2012,Sentef2013}. 

\begin{figure}
\centering\includegraphics[width=1\columnwidth]{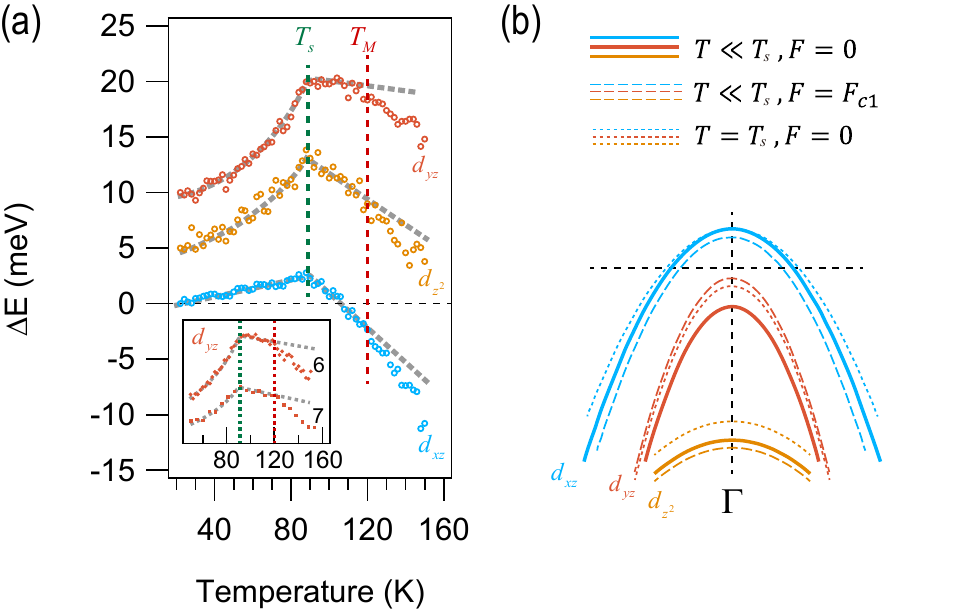}
\caption{
Comparison between equilibrium thermally and non-equilibrium optically pump-induced phase transitions.
(a) Equilibrium-temperature-dependent energy shift of $d_{xz}$, $d_{yz}$, and $d_{z^2}$ bands. $T_s$ denotes temperature of nematic phase and structural transition; $T_M$ denotes temperature of rising band gap between hole and electron bands at point M. The momentum in the analysis of each band are the same as that in Fig.~\ref{Fig2}.
The inset shows additional temperature dependent measurements of the $d_{yz}$ band on the other two samples with different photon energies and different momentum cuts. For the 6 eV (p-polarized) probe, the cut is along X, and for the 7 eV probe, the cut is along Y (Supplemental Discussion \#3\cite{SM}).
(b) Schematic showing the difference between equilibrium thermally induced and optically pump-induced band shifts.
}
\label{Fig4}
\end{figure}

To reveal the underlying physics of the two ultrafast electronic phase transitions, detailed equilibrium-temperature-dependent measurements were taken at the Brillouin-zone center (Fig.~\ref{Fig4}(a)). The temperature-dependent band shift of \dyz~is similar to the fluence-dependent band shift, as evidenced by two temperature scales at approximately 90 K ($T_s$) and 120 K ($T_M$).
The consistency of the shape of the curve of the \dyz~band shift between the temperature- and pump-fluence-dependent measurements, and the similar energy scale between that from the non-equilibrium quasiparticle rate and the band gap at point M below 120 K strongly suggest that \fct~is the critical fluence quenching the M point \dyz-\dxz~band gap (the orbital order) and that \fco~is the fluence draining the nematic states (Supplemental Discussion \#2\cite{SM}). It is consistent with the estimated non-equilibrium electronic temperature at 0.2 ps for the pump fluence of \fco~and \fct, is approximately at $T_s$ and $T_M$, respectively (Supplemental FIG. 1 and Discussion \#1\cite{SM}).

$T_s$ can be also clearly evidenced in the temperature-dependent band shifts of the \dxz~and \dz~orbits. It is interesting that the bands \dxz~and \dz~shift to higher binding energies below $T_s$, opposite to the case in which both of the bands shift to lower binding energies below the \fco~in Fig.~\ref{Fig2}. The absence of $T_s$ in the fluence dependent band shifts of the \dxz~and \dz~orbits suggests that ultrafast structural transition does not happen in the studied time and fluence range, agreeing with the fact that the timescale of the ultrafast photon-excitation-induced lattice distortion transition at similar photon energy is much longer or needs a much higher pump fluence in iron-based superconductors \cite{Patz2014,Rettig2016}. The absence of structural transition indicates that the electronic states can be isolated from the lattice shortly after ultrafast photoexcitation, and the above-observed critical pump fluences at \fco~and \fct~are purely electronic origin. From above, it can be concluded that the structural transition at the $T_s$ is driven by the nematicity. Since the orbital transition temperature is higher, we can conclude that the nematicity, which is also a broken of rotational symmetry, is driven by the orbital order.

The difference between the optically pump-induced energy-band shift and temperature-dependent band shift near $\Gamma$ is summarized in Fig.~\ref{Fig4}(b).
Ultrafast photoexcitation strongly modulates the energy band near the Fermi energy and drives the up-shift of the \dyz~bands and down-shift of the \dxz~and \dz~ bands below the pump fluence of \fct. Equilibrium thermal excitation induces similar energy-band shifts for the \dyz~orbit, while in contrast, the \dxz~and \dz~bands shift opposite to that below \fco. 
All the bands near the Fermi energy shift down upon increasing the pump fluence above \fct~or heating the sample above $T_M$ synchronously. Above \fct~or $T_M$, the band evolution is possibly a result of the chemical potential shift since there is a great enhancement of the density of states near the Fermi energy due to the closure of the M-point band gap, consistent with a previous report \cite{Rhodes2017}. The monotonic down-shifts of the \dxz~and \dz~bands may also be the results of the chemical potential shift after photoexcitation.

As schematized in Fig.~\ref{Fig4}(b), the different behavior between the photoexcited energy-band shifts in Figs.~\ref{Fig2}(d)--(f) and temperature-induced band shifts in Fig.~\ref{Fig4}(a) suggests that some additional phase transition not sensitive to photoexcitation, such as the structural transition, must be counted to the electronic transition at $T_s$. By subtracting the photoexcitation-driving band shifts (purely electronic order) from the thermal-driving ones (purely electronic order + additional phase transition), the estimated net energy shifts for the \dyz~, \dxz, and \dz~orbits from the additional transition in the temperature dependent measurement are approximately -1, 5, and 12 meV, respectively. Such additional band shifts could be a result of different chemical potential shifts between the photoexcitation and equilibrium heating, but can be ruled out by the fact that the \dyz~band behaves similarly in the two experiments. Any order that is also not sensitive to photoexcitation may explain our data, but no other long-range order has been evidenced in FeSe to date \cite{Boehmer2017}. Magnetic fluctuations at the same temperature scale of the structural transition were observed in FeSe \cite{Wang2016}, but it is also less possible, since magnetic fluctuations can be also destroyed by ultrafast photoexcitation \cite{Patz2014}. The only additional order that is not sensitive to photoexcitation at such low pump fluences and short delay times is the structural transition. The observed additional band shifts in the \dxz~and \dz~bands are even larger than the purely electronic band shift of the \dyz~orbit, indicating that the structural transition plays an important role in the nematic electronic phase. However, it is quite anomalous that for a lattice constant change of less than 0.5\%~in FeSe, the estimated structurally induced band shift from conventional theory is 1 order smaller \cite{McQueen2009,Boehmer2013,Baek2015,Gerber2017}. Our findings may suggest that there is possible strong lattice-nematicity coupling in the iron-based superconductor. Further experimental and theoretical studies are necessary to clarify the underlying mechanism of the anomalous contribution to the electronic order from the structural transition.

In summary, high-resolution trARPES measurements were conducted on FeSe superconductors, and  two critical pump fluences were found that correspond to two purely ultrafast electronic phase transitions, appearing only in the photoinduced ultrafast band shifts of the \dyz-orbit-derived band within our time and energy resolution. By comparing the ultrafast photoexcited and equilibrium-temperature-induced band energy-shifts, we found an anomalously large contribution to the nematic electronic states from the structural transition. 

\begin{acknowledgments}
W.T.Z. acknowledges support from the National Key R\&D Program of China (Grant No. 2021YFA1401800 and 2021YFA1400202) and National Natural Science Foundation of China (Grant No. 11974243 and 12141404) and additional support from a Shanghai talent program. D.X. acknowledges support from the National Natural Science Foundation of China (Grant No. 11925505). D.Q. acknowledges support from the National Natural Science Foundation of China (Grant No. 12074248). The work at Fudan University was supported by the National Natural Science Foundation of China (Grant No. 11874119), the Innovation Program of Shanghai Municipal Education Commission (Grant No. 2017-01-07-00-07-E00018) and the Shanghai Municipal Science and Technology Major Project (Grant No. 2019SHZDZX01).

\end{acknowledgments}

\bibliographystyle{apsrev4-2}
%
\newpage
\begin{figure*}
\includegraphics[width=2\columnwidth,page=1]{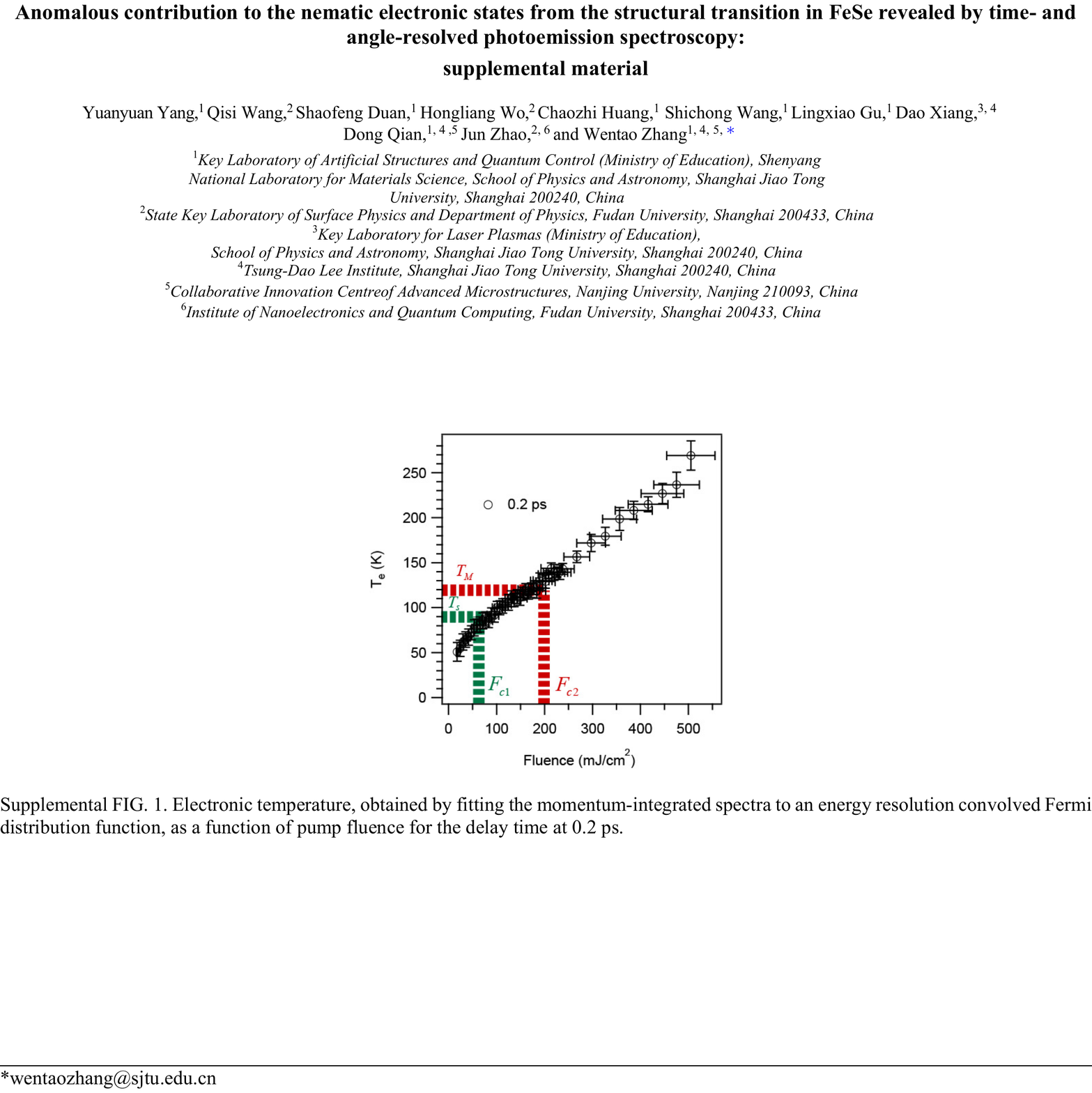}
\end{figure*}
\begin{figure*}
\includegraphics[width=2\columnwidth,page=2]{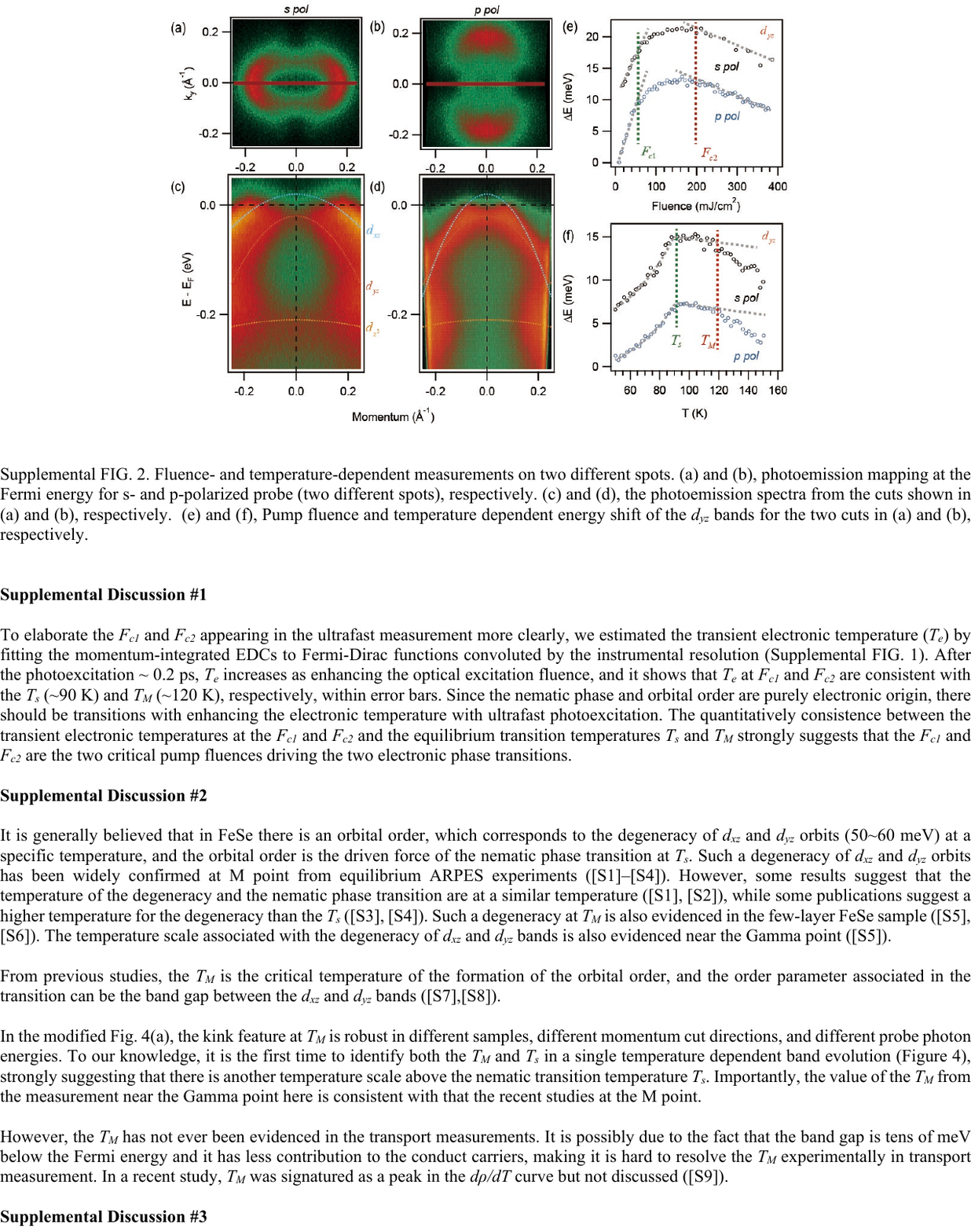}
\end{figure*}
\begin{figure*}
\includegraphics[width=2\columnwidth,page=3]{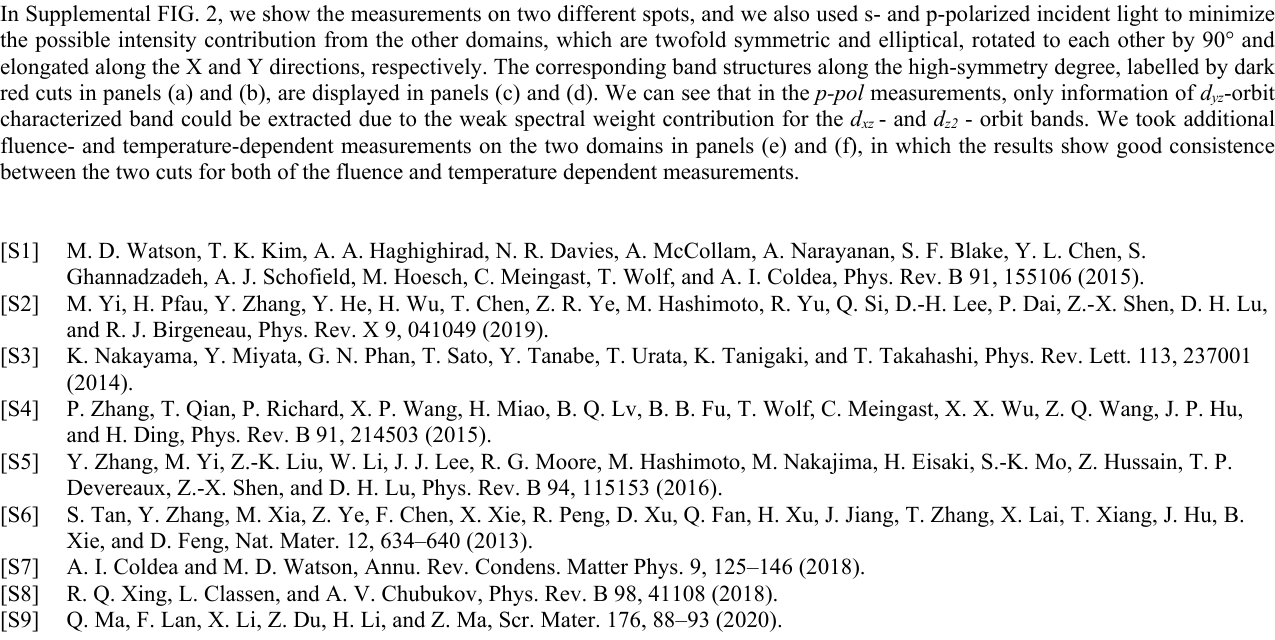}
\end{figure*}

\end{document}